# Future of Flexible Robotic Endoscopy Systems

Tian En Timothy Seah[1], Thanh Nho Do[1,a], Nobuyoshi Takeshita[2], Khek Yu Ho[2], and Soo Jay Phee[1]

[1]School of Mechanical and Aerospace Engineering, Nanyang Technological University, 50 Nanyang Avenue, Singapore 639798
[2]Department of Medicine, Yong Loo Lin School of Medicine, National University of Singapore and National University of Health System, Singapore 119260, Singapore

*Corresponding author*: [a]thanh4@e.ntu.edu.sg

## Abstract

Robotics enables a variety of unconventional actuation strategies to be used for endoscopes, resulting in reduced trauma to the GI tract. For transmission of force to distally mounted endoscopic instruments, robotically actuated tendon sheath mechanisms are the current state of the art. Robotics in surgical endoscopy enables an ergonomic mapping of the surgeon movements to remotely controlled slave arms, facilitating tissue manipulation. The learning curve for difficult procedures such as endoscopic submucosal dissection and full-thickness resection can be significantly reduced. Improved surgical outcomes are also observed from clinical and pre-clinical trials. The technology behind master-slave surgical robotics will continue to mature, with the addition of position and force sensors enabling better control and tactile feedback. More robotic assisted GI luminal and NOTES surgeries are expected to be conducted in future, and gastroenterologists will have a key collaborative role to play.

## INTRODUCTION

Computer and robotically-assisted surgery was developed to overcome the limitations of minimally-invasive surgical procedures, as well as enhance the interventional capabilities of surgeons. The poster child for robotic surgery in the 2000s was the Da Vinci machine, which enhanced the visualization and manipulation abilities of laparoscopic surgeons. Originally developed for battlefield telesurgery by DARPA, it makes use of a master-slave configuration, in which the user sits at a master control terminal and remotely operates a slave robot. Other medical specialties have also embraced robotic technology to improve precision, safety and reliability. For example, ophthalmic surgery benefits greatly from handheld devices which filter out tremors from the surgeon's hand [1]. Neurosurgery requires thorough preoperative imaging and path planning which is then put into action by a flexible path following robot. Endovascular surgery is also heavily imaging based and benefits from the use of master-slave control interfaces to increase ergonomics and reduce fluoroscopy exposure to the surgeon [2].

In gastroenterology, robotics has been implemented in flexible endoscopes to increase their effectiveness and safety, as well as to augment their therapeutic capabilities. Often, haustral folds create obstructions that the tip mounted video camera cannot see behind. Robotic mechanisms have been added to the heads of endoscopes that let them bend 180° backwards to see behind the folds. Novel locomotion methods that reduce the amount of force applied to the intestinal walls and interventional endoscopes have also been developed. In the most common embodiment, instruments emerge from the tip of the endoscope into the camera's field of vision. These instruments are similar to those used in laparoscopic surgery, such as tissue graspers, electrocautery devices, and wire loops. They are usually cable actuated and controlled from the proximal end by the endoscopist. Endoscopes with two or more instrument channels allow bimanual grasping of tissue and/or resection ability. They are also robotically-assisted, which offers an improvement to conventional surgical endoscopy by addressing some of the inherent challenges of providing enough force to distally mounted tools through a long and flexible conduit. The most promising flexible endoscopic robots being developed will be introduced further in this chapter and a case series on endoscopic submucosal dissection (ESD) using one of the platforms will be presented.

Finally, Natural Orifice Transluminal Endoscopic Surgery (NOTES) is a new paradigm that makes use of natural orifices to access the peritoneum for surgery, thus leaving no visible scars. Current technology limits it to transoral, transvaginal, and transanal avenues of access, but with miniaturization even more may be possible. Much of the technology for NOTES is based on that of existing gastrointestinal surgery. Within this context, the future developments and possibilities of flexible robotic endoscopes will be discussed.

## ROBOTIC ENDOSCOPY SYSTEMS

The current robotic endoscope platforms being developed are summarized in Table 1. Information is provided on their key features (degrees of freedom, dimensions) and readiness to market (clinical trials, regulatory approval) [3, 4]. Direct comparisons are not useful due to the differing areas of anatomy that they specialise in, for example, oropharyngeal vs colonic. Where they do operate in the same anatomical domain, at present there is insufficient clinical data to suggest the superiority of one device over another.

The diagnostic endoscopes presented employ unique methods to advance through the colon that reduce the amount of force exerted on the colon, reducing trauma to the endoluminal lining. These would be cumbersome to operate using a manual approach, so automation allows the endoscopist to focus on the task of identifying features of interest such as polyps and lesions.

The surgical endoscopes presented also benefit greatly from robotic assistance. Most surgical endoscopes make use of control tendons actuated at the proximal end to move distally mounted instruments. The convoluted path (tortuosity) of an endoscope through the lower GI tract and the bending of the endoscope tip into a retroflexed position (i.e. 180°) introduce friction between the tendons and their guide sheaths [5-11]. This friction leads to a sluggish response of the end effector to human commands made at the proximal end and a jerky start-stop motion of the surgical tools. These control problems can be mitigated by the introduction of a robotic controller which intelligently compensates for nonlinear and hysteretic friction effects [12-16]. Friction losses can also severely degrade the output force, so to maintain the payload of the end effectors for tissue manipulation, motorised actuation is desirable [10, 16-18].

In addition, better ergonomics are possible with robotics. The control system of manually operated endoscopes has been modelled after that of laparoscopic tools, which make use of control wheels and dials to directly control the tendons. This is unintuitive and results in a steep learning curve. As with the da Vinci machine, robotics offers a more natural mapping of user motion to the movement of the surgical tools. Haptic feedback can also be incorporated, which improves the precision at which a surgeon can manipulate the tissue.

Table 1: Current robotic endoscopy platforms

| Name | Flexible Length (cm) | Diameter (mm) | Number of Instrument Channels | Regulatory Approval | Commercial status | Animal Trials | Human Trials |
|---|---|---|---|---|---|---|---|
| **Aer-O-Scope** | >150 | - | - | FDA, CE | Under development | Yes | Yes |
| **Endotics** | ~200 | 17 | - | CE | Available | Yes | Yes |
| **Invendoscope** | 200 | - | 1 | FDA, CE | Under development | Yes | Yes |
| **Neoguide** | 173 | 14-20 | - | FDA | Under development | Yes | Yes |
| **Flex** | ~50 | Variable | 2 | FDA, CE | only in Europe | Yes | Yes |
| **i-Snake** | 20 | 14 | 1 | - | - | Yes | No |
| **MASTER** | 154 | 22 | 2 | - | Under development | Yes | Yes |
| **Viacath** | 90 | 16 | 2 | - | Available | Yes | - |

# Diagnostic Endoscopes

**1. Aer-O-scope**

The Aer-O-scope is equipped with a front facing camera and a 360° panoramic camera for viewing the side walls of the colon (see Fig. 1 and Fig. 2). This unique optical setup allows it to see behind haustral folds. Upon insertion, a balloon at the anal sphincter is used to make the colon airtight. Two balloons mounted at the tip of the scope are inflated with $CO_2$ gas to provide cushioning as it slides through the colon. Forward motion is achieved by pressurising the segment of colon between the proximal balloon and the middle balloon [19].

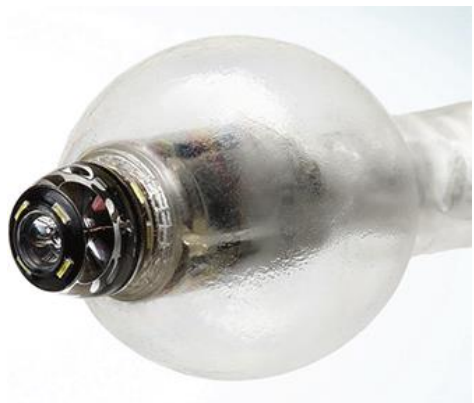

Fig 1: Omni-view cameras on tip of endoscope with inflated distal balloon. Source [giview.com]

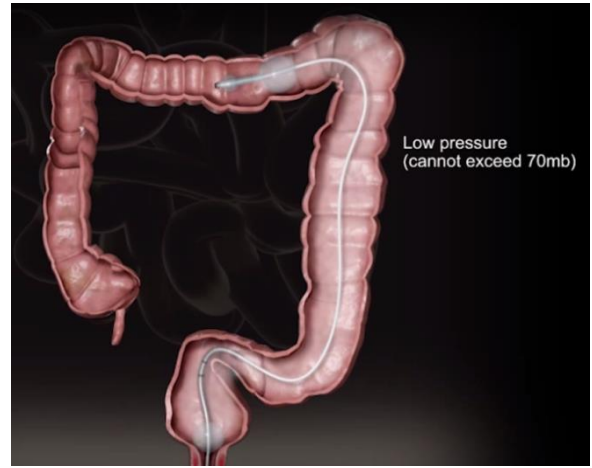

Fig 2: Pneumatic intubation

## 2. Endotics

The Endotics endoscope draws inspiration from the locomotive principle of an inchworm. It has two anchoring points, proximal and distal, which are alternately actuated using a vacuum suction mechanism. The middle section is able to contract and expand along its longitudinal axis (See Fig. 3 and Fig. 4). By cycling between the 4 actuation mechanisms, it is capable of inching forwards or backwards [20].

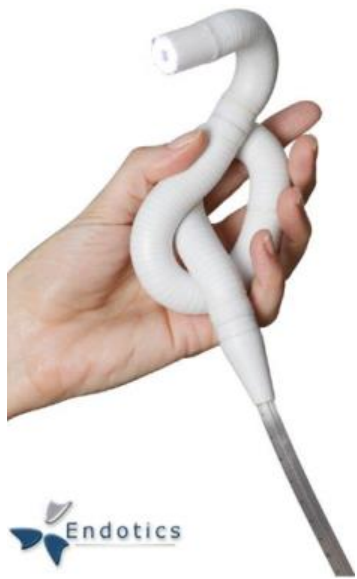

Fig 3: Endotics endoscope [www.intechopen.com]

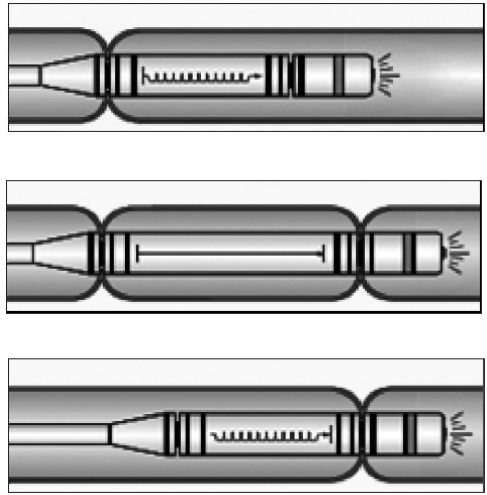

Fig 4: Endotics principle of locomotion

## 3. Invendoscope

The Invendoscope aims to be a lightweight, single use colonoscope that addresses the medical risk of cross contamination from improperly sterilized endoscopes. It has a robotic hydraulically articulated tip for navigation and retrograde viewing. It uses a novel way of advancing through the colon, being surrounded by an air filled inverted sleeve that cushions the lumen as the tip moves forward (see Fig. 5). This theoretically reduces the amount of force exerted on the colon [21]. In a comparison study between the Endotics and Invendoscope systems, it was shown to be faster at completing the colonoscopy, but induced greater levels of discomfort [22].

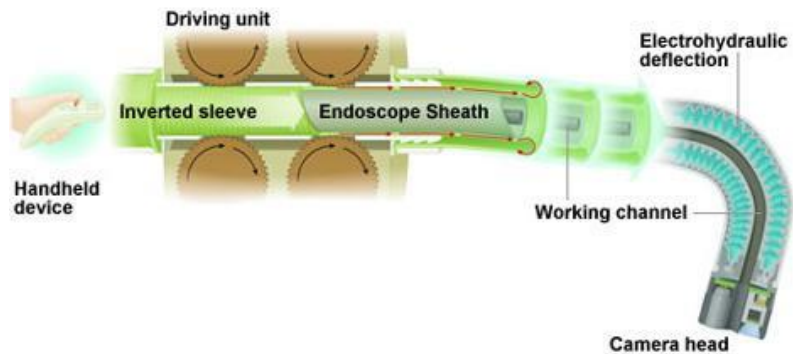

Fig 5: Invendoscope operating principle. [www. medgadget.com]

## 4. Neoguide

The Neoguide endoscopy system, recently acquired by Intuitive Surgical, consists of 16 individual segments that can be programmed to change their shape (see Fig. 6). As the lead segment makes its way deeper into the colon, the rest will automatically change their shape to follow the path that has been defined. Each of the 16 segments has 2 degrees of articulation [23].

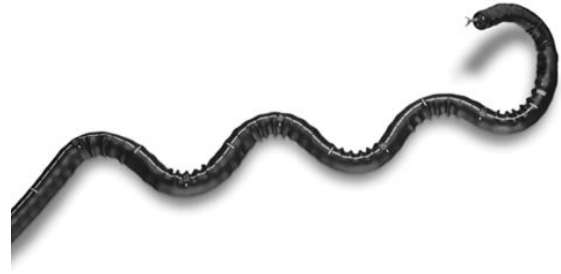

Fig 6: Neoguide endoscope

# Surgical Endoscopes

**1. Flex robot**

Medrobotics' Flex Robotic System, the product of research at Carnegie Mellon University, is a short flexible endoscope designed for transoral surgery (See Fig. 7 and Fig. 8). While transoral pathologies do not normally fall under the scope of gastrointestinal specialists, it has operating principles common to endoscopes. It has an overtube comprising many ball and cup vertebral segments. Articulation is achieved by motorized tensioning of a cable tendon system that runs through the overtube, which also houses a video camera at the distal tip. Due to an inner spine that can be stiffened, it is able to maintain the shape of the path travelled. Riding on the outside of the overtube are two instrument channels that allow various manual cable instruments to be inserted and controlled, such as graspers and monopolar cutters [24, 25].

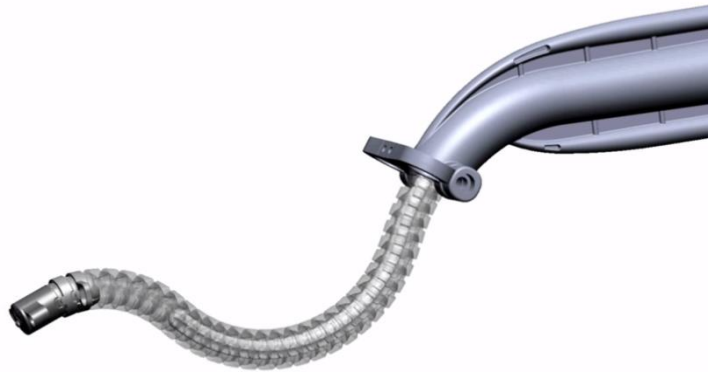

Fig 7: Medrobotics Flex with translucent overtube for illustration. [medrobotics.com]

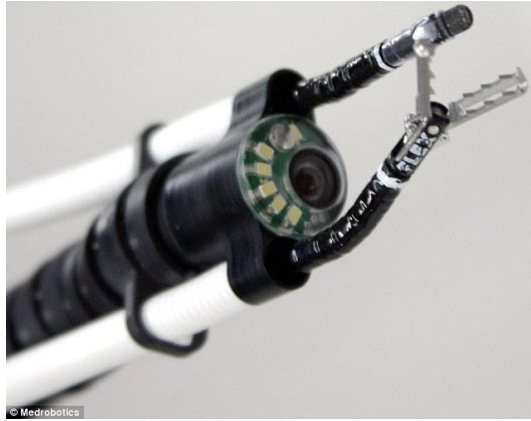

Fig 8: Graspers and video camera on Medrobotics Flex

**2. i-Snake**

The i-Snake, being researched at Imperial College London, is a short flexible robotic tool [26]. It has 6 segments linked by movable joints (see Fig. 9 and Fig. 10). Micromotors embedded within the segments enable it to move in 7 degrees of freedom by using miniature gears and pulleys. A video camera is mounted at the tip and it also has an internal instrument channel for endoscopic tools. At present, the large size of the motor segments hinders the minimum radius of curvature that it can achieve, thus limiting it to peritoneal surgery.

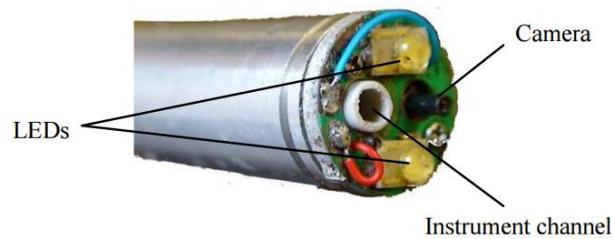

Fig 9: Tip of the i-Snake.

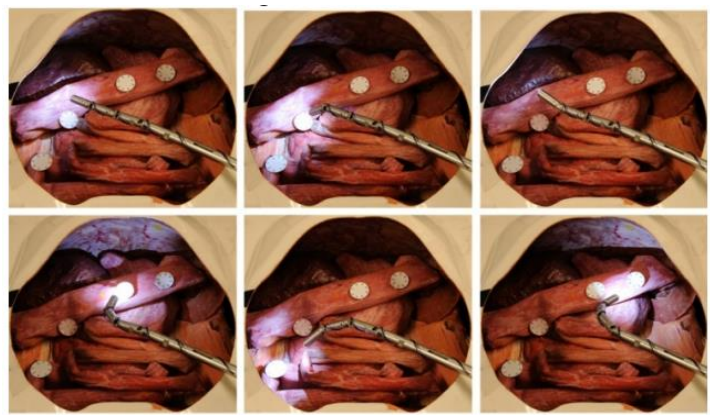

Fig 10: Articulation of i-Snake.

**3. MASTER**

The Master and Slave TransEndoluminal Robot (MASTER), developed at Nanyang Technological University and commercialized under the company Endomaster, consists of a custom developed endoscope that has an integrated videoscope and 2 instrument channels [27]. The instrument channels support a variety of instruments with cable articulation, such as electrocautery tools and 7 degree of freedom graspers (see Fig. 11). Clinical demonstrations have been performed, such as NOTES liver resection and endoscopic submucosal dissection (ESD) [28]. However, questions remain about sterilization procedures as it has not yet been granted FDA or CE Mark approval. Detailed descriptions of ESD, full thickness gastric wall resection, and hepatic wedge resection procedures will be given in the next section.

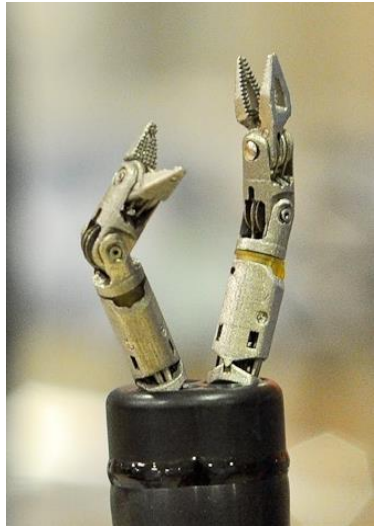

Fig 11: Twin 7 DOF arms on Endomaster's MASTER robot. [endomastermedical.com]

### 4. Viacath

The Viacath (Hansen Medical systems, USA), a flexible cable robot initially developed for endovascular and urological interventions, can also be used in the field of gasteroenterology (see Fig. 12). It consists of a steerable overtube that houses a standard endoscope and two instrument channels. It shares its control interface and actuation mechanism with the Laprotek robotic assisted laparoscopy system, also marketed by the same company. This facilitates tool changes. Initially it could only exert a weak tip force of 0.5N, but the latest version reportedly can exert up to 3N [23].

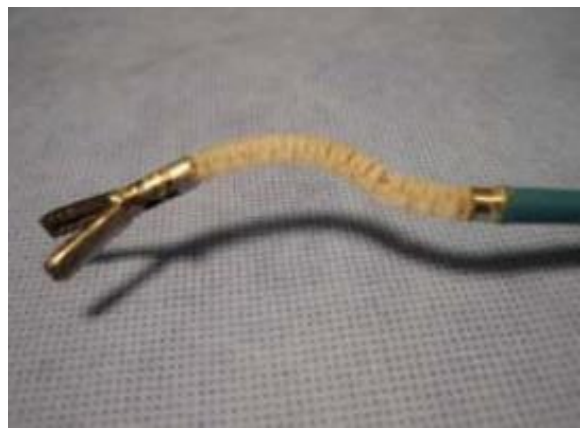

Fig 12: Viacath 6 DOF robotic arm

# ENDOSCOPIC SUBMUCOSAL DISSECTION WITH MASTER ROBOT

Endoscopic mucosal resection (EMR) is an endoscopic technique in which specimens of early stage superficial lesions are resected. Most early stage GI cancers are confined to the mucosal and submucosal layers and have not yet progressed to deep submucosal invasion or lymph node metastasis. Thus, early detection and removal lead to a high chance of patient survival; in contrast, the survival rate for advanced gastrointestinal cancers remains poor. It is also known that en bloc resection reduces the risk of residual cancer.

Endoscopic submucosal dissection (ESD) was developed as an improvement to EMR, and provides the ability to resect larger lesions en bloc and with greater margins. It is increasingly recognised as a highly effective procedure for the treatment of early stage gastric cancers. Compared with EMR, it reduces the rate of local recurrence from 15% to 1% and allows more accurate histological examination of the resected specimen [29]. However, en bloc dissection along the submucosal layer is difficult due to the technical limitations of current therapeutic endoscopes, which are equipped with poorly manoeuvrable cutting tools. The risk of procedural complications such as perforation and delayed bleeding means that ESD is performed by only the most skilled of endoscopy surgeons. A long procedure time adversely affects patient recovery after ESD procedures, and may lead to ulceration and other negative effects [30]. A study conducted at Hiroshima University Hospital across 896 patients found that the incidence of intraoperative bleeding with ESD was 22.6% compared with 7.6% for EMR. Perforation was significantly higher at 53.8% with ESD compared with 2.9% for EMR [31]. Clearly, there is significant scope to reduce the incidence of trauma to the patient during ESD. The use of a robotic master-slave interface to carry out the procedure has been shown to reduce operating times and may enable novice surgeons to perform a satisfactory job.

## Indications

ESD can be considered for the removal of superficial premalignant and well to moderately differentiated malignant lesions in the GI tract. The Japanese Society of Gastrointestinal Endoscopists categorises superficial (Type 0) lesions into polypoid and nonpolypoid categories, which are further subcategorised as shown in Fig 13. The lateral extent of the lesion can be detected by methylene blue dye spraying and advanced imaging modalities such as narrow band imaging.

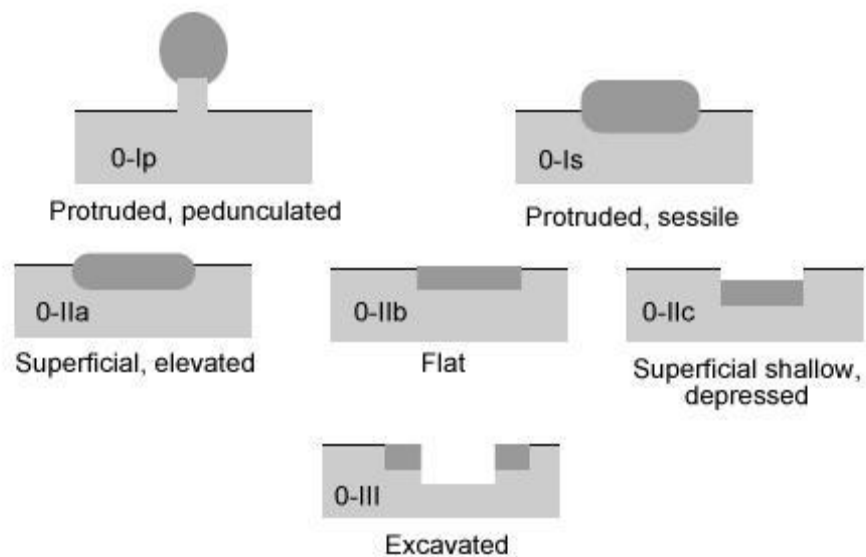

Fig 13: Paris classification of superficial gastrointestinal neoplasms [32]

The depth of the invasion of the lesion can be determined by high frequency endoscopic ultrasound (EUS), which produces a stratified image of nine separate layers (Fig 14). The submucosa is divided into 3 layers, sm1, sm2, and sm3. The images produced can help in estimating the risk of lymph node metastases and thus the suitability of ESD in resecting the lesion.

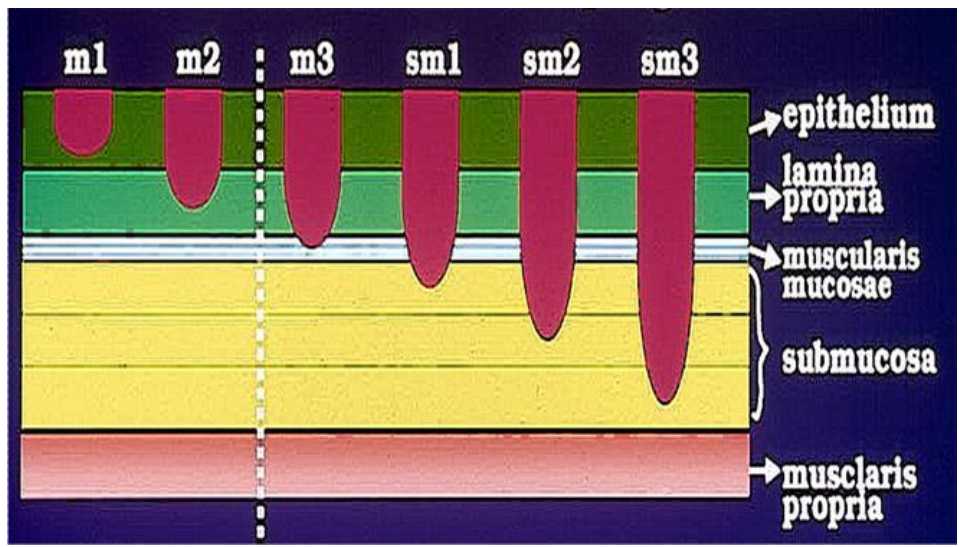

Fig 14: Depth of lesion penetration as determined by high frequency endoscopic ultrasound [33]

Earlier techniques like EMR are limited to small (<20mm) lesions, or piecemeal resections of larger lesions. In contrast, ESD allows for the resection of large ulcerative lesions greater than 20 mm in diameter. High en bloc resection rates of 80-90% can be achieved with ESD compared with 50% for EMR [29]. This increases the accuracy of histopathological assessments, which in turn helps physicians to determine the best course of management for patients.

## Contra-Indications

ESD is considered unsuitable when there is a suspicion of lymph node metastasis or deep tissue invasion by the lesion. In addition to EUS, submucosal invasion can also be detected by the injection of a saline solution into the submucosa underneath the lesion. This is known as the non-lifting test. If the lesion fails to elevate above the surrounding tissue, it is considered a positive test result. In this scenario, full thickness gastrotomy resection or other more extensive methods are recommended.

The technical complexity of the operation leads to long procedure times, (90 minutes compared with 30 minutes for EMR) hence it is contraindicated for patients with possible complications from premedication or who are susceptible to surgical stress. Factors increasing the difficulty of the operation include scarring on or around the lesion, which indicates a thin submucosal cushion and hence increased probability of perforation through the muscularis propria. The upper part of the stomach is also challenging due to its large vascular network, which can cause attempts at haemostasis to fail [34]. Because of these difficulties, ESD is generally not recommended for novice surgeons [35].

## Instruments and Accessories

The 1st generation of the MASTER system consists of 3 major components: a master controller, a telesurgical workstation, and a pair of slave arms equipped with a grasper and a monopolar electrocautery hook. The slave arms access the surgical site through the operating channels of a conventional forward-viewing therapeutic endoscope. The operation of MASTER robot is depicted in Fig. 16. The master controller has receptacles for the surgeon's right index finger and thumb and a handle for his/her left arm to grasp. These control the gripping action of the grasper and cutting motion of the monopolar diathermy "L" hook respectively. The current to the diathermy device is activated by a foot pedal. The surgeon's hand motions are recorded by the robot arms using robotic 'proprioception'. These are converted by the master console into control signals, which are sent by data cable uplink to the slave manipulators. The slave motors then tension/slacken the tendons by appropriate amounts to mimic the user's hand motions on the slave robot. Each motor-tendon pair operates 1 degree of freedom of the slave arms. The surgeon is assisted by an endoscopist, who controls the macro-level positioning and orientation of the endoscope. Visual feedback to the surgical team is provided by the camera on the endoscope.

More recent iterations of the MASTER system feature interchangeable instruments such as injection needles, grasping forceps, and various cautery devices. A version of electrocautery device that was developed for EMR is the insulated tip knife, which has a small ceramic ball attached to the tip of a cutting needle (Fig. 15). The ceramic ball restricts cutting to the lateral plane, preventing depthwise perforation of the muscularis propria.

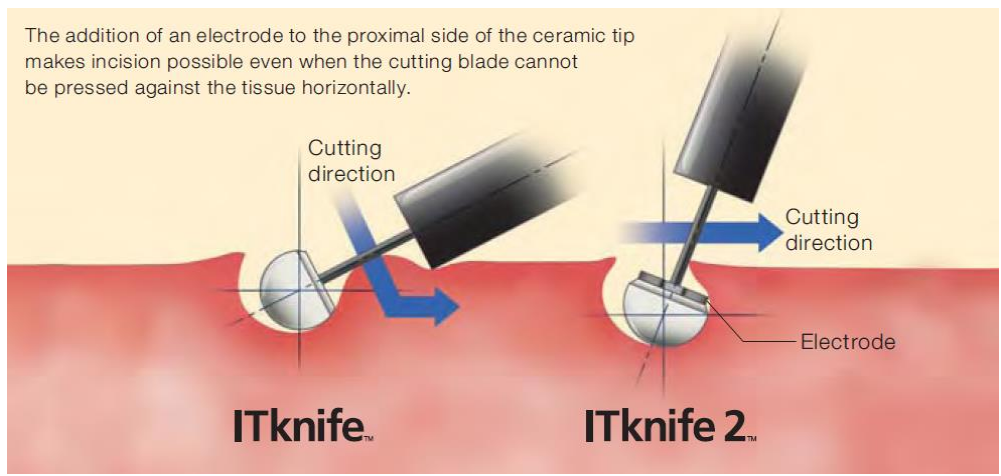

Fig 15: Olympus insulated tip knives [36]

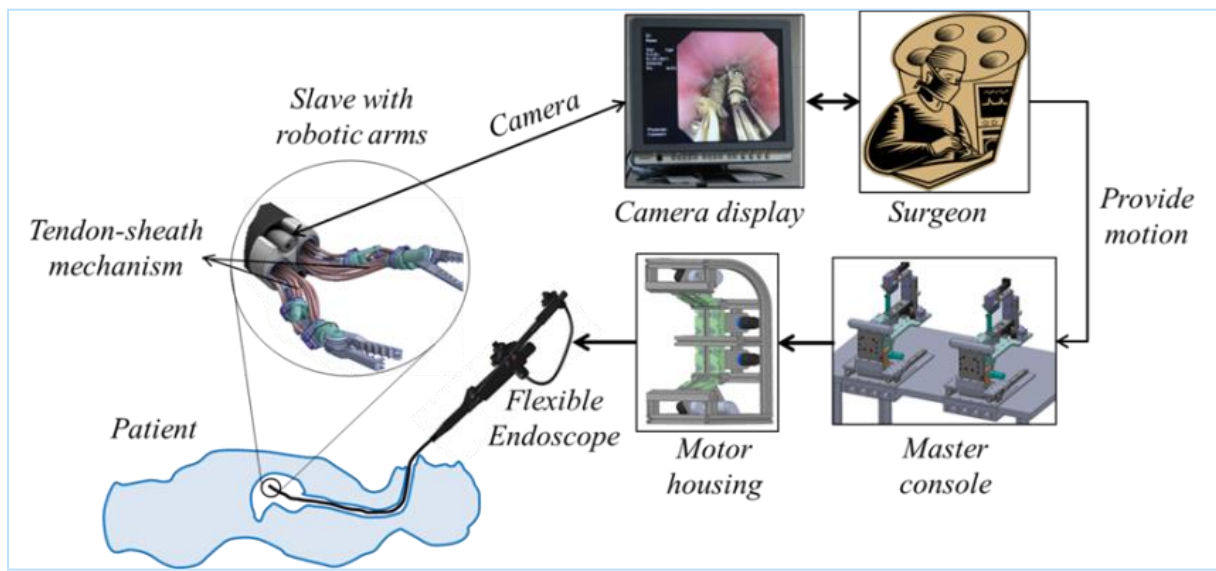

Fig 16: Flow of operation for the MASTER robot

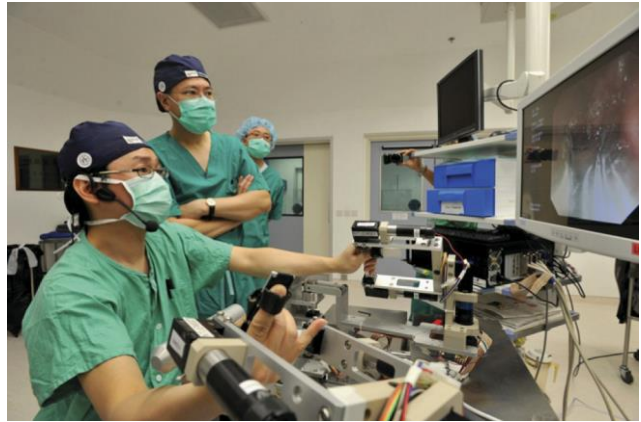

Fig 17: Clinical setup for the performance of robotic ESD, with 1 endoscopist holding the endoscope while the surgeon performs sub-mucosal dissection using the robotic arms.

## Pre-procedure Preparation

The endoscope and the end effectors are sterilized by immersion in glutaraldehyde for 30 minutes. To clear the gastric cavity, patients have to fast for a minimum of 6 hours before the procedure (fluids are allowed). They are sedated under general anaesthesia and ventilated via naso or orotracheal intubation. They are laid sideways in the left lateral position, which makes blood in the stomach gravitate towards the fundus and greater curvature.

## Master-assisted ESD Procedure

The ESD procedure consists of two main parts: marking of the lesion, and submucosal dissection. The endoscopist first introduces a standard therapeutic videogastroscope into the patient and steers it to the surgical site (See Fig. 17). The lesion is circumferentially marked with electrocautery. Then, fluid (mixture of 100 mL normal saline, 5 mL of indigocarmine / methylene blue and 1 mL of 1 in 10,000 epinephrine and sodium hyaluronate) is injected into the submucosa to elevate the lesion. The actual amount used depends on the size of the lesion and swelling response to the fluid. A lack of elevation response may indicate that the lesion has deeply invaded the submucosa or metastasised into the lymph nodes, in which case ESD would not be able to excise the entire lesion but could still offer useful histopathology information. Next, a small incision is made at the distal end of the lesion using a needle knife or dual knife. The remaining circumference of the lesion is then scribed with an insulated tip knife.

The conventional endoscope is removed and the MASTER-equipped endoscope is inserted to conduct the robotic submucosal dissection. For the 1$^{st}$ Generation MASTER, the grasper arms are unable to fully retreat into the endoscope so an overtube must be used during its introduction to prevent damage to the tracheal lining. The grasping arm is used to grasp and retract the tumour-side open edge of the mucosa in order to expose the submucosa. The cautery arm is then able to perform dissection of the submucosa.

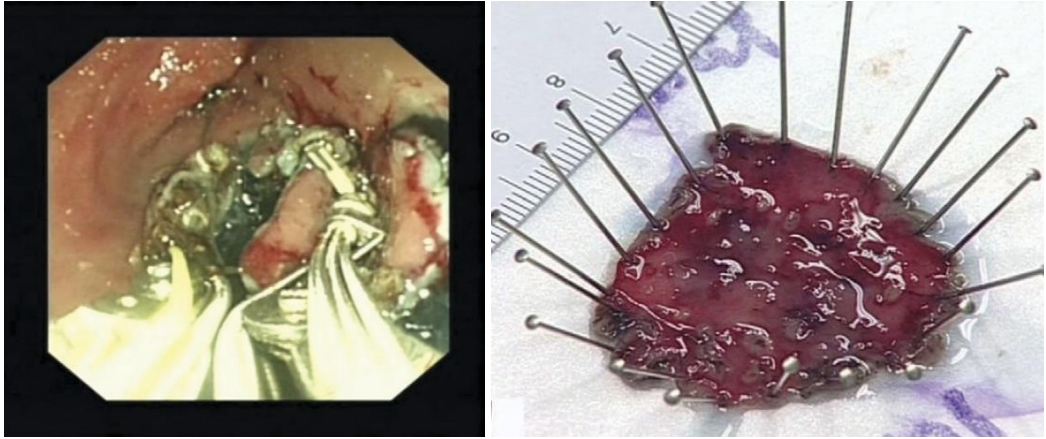

Fig 18. ESD result: (Left) The procedure of ESD by MASTER with adequate retraction to demonstrate the submucosal plane for dissection; (Right) Specimen after MASTER ESD

## Post ESD Management

After the procedure, the patient's vitals such as blood pressure, pulse, and arterial oxygen content are monitored hourly. They are prescribed with a high dose of proton pump inhibitor to limit the production of gastric acid, thus reducing the occurrence of strictures. Follow up endoscopy is performed 3 months later to confirm that the ESD-induced ulcer has healed and that the tumour has not recurred.

## Results of Clinical Trials

A multicenter prospective endoscopic submucosal dissection (ESD) study has been performed [27]. Five patients with a diagnosis of early-stage gastric neoplasia, limited to the mucosa were recruited from 2 centers. The results are given in Table 2.

Table 2: Procedure timing in minutes for MASTER ESD

|  | **Patient 1** | **Patient 2** | **Patient 3** | **Patient 4** | **Patient 5** |
|---|---|---|---|---|---|
| Center | India | India | India | Hong Kong | Hong Kong |
| Lesion marking | 3 | 4 | 4 | 2 | 1 |
| Submucosal injection | 3 | 2 | 2 | 3 | 3 |
| Circumferential mucosal incision | 4 | 18 | 15 | 5 | 7 |
| Insertion of overtube | 4 | 4 | 4 | 5 | 5 |
| Exchange of the endoscope | 3 | 3 | 3 | 3 | 3 |
| Robotic submucosal dissection | 19 | 5 | 3 | 50 | 16 |
| Total procedure time | 26 | 36 | 31 | 68 | 35 |

All patients underwent successful MASTER-assisted ESD (See Fig. 18). The mean submucosal dissection time was 18.6 minutes (median, 16 minutes; range, 3-50 minutes). No perioperative complications were encountered. All patients were discharged from the hospital within 3 days after procedures. Two patients were found to have intramucosal adenocarcinoma, 1 had high-grade dysplasia, 1 had low-grade dysplasia, and 1 had a hyperplastic polyp. The resection margins were clear of tumors in all 5 patients. No complications were observed at the 30-day follow-up examination (Fig. 19). Follow-up endoscopic examinations revealed that none of the patients had residual or recurrent tumors.

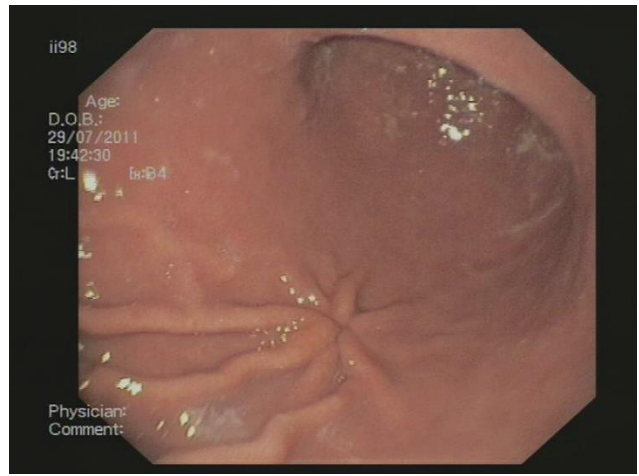

Fig 19. Thirty-day postoperative endoscopic picture

# FULL THICKNESS GASTRIC RESECTION WITH MASTER ROBOT

Gastrointestinal stromal tumours (GIST) originate from the interstitial cells of Cajal in the connective tissue, or stroma, of the stomach, rather than the mucosal lining. Because they occur on the muscularis propria, mucosal and submucosal dissections are ineffective forms of treatment. Laparoscopic gastric wedge resection has become the gold standard for removal of GIST, being a short procedure with reduced trauma that allows for next-day discharge of the patient. In this surgical procedure, access to the resection site is obtained through the abdomen. The desired use of robotic endoscopy to carry out the resection encounters several challenges. Firstly, insufflation of the GI tract is lost when the resection is carried out. Secondly, the submucosal techniques from ESD provide insufficient tissue retraction and exposure. Lastly, the luminal defect created by the dissection cannot be closed by endoclipping, and the first generation of the MASTER robot provides only 1 grasper which is insufficient for suturing. To overcome these barriers, a clinical approach has been developed that uses the MASTER as a platform for retraction and dissection, and the Apollo Overstitch as a device for tissue approximation [37].

## Indications

GIST is a rare form of cancer and comprises only 1 percent of GI tumours. However, almost one third of GIST masses are malignant or at high risk of malignance. Adjuvant therapies are ineffective, leading to high mortality rates. Masses can be classed as having metastatic potential based on their size and on the histological analysis of tissue samples obtained by diagnostic endoscopy. In one study, more than 5 mitoses per fifty high powered fields and having a maximum diameter greater than 10 cm indicated an 86% chance of eventual metastasis [38]. In contrast, being below the threshold for both indicators led to only a 2-3% chance of metastasis and so in these cases the tumour may be regularly monitored instead. 5 cm in diameter has been suggested as the predictive threshold for malignancy and hence surgical treatment.

## Procedure

The procedure consists of four main parts. First, using a single channel endoscope, the anterior wall of the stomach is slung to the abdominal wall and affixed using a Loop Fixture II device. Second, the gastric lesion is circumferentially marked using a needle knife. A mixture of saline and indigocarmine/methylene blue is injected to elevate the lesion. A mucosal incision is made at a point on the circumference with an IT knife and a needle knife. Third, the endoscope is withdrawn and the MASTER is introduced via an overtube. The mucosal incision is completed around the lesion to expose the muscularis propria, which is then grasped and incised to the serosa. The full thickness resection is completed using retraction provided by the grasper and dissection with the electrocautery hook. At this point, a loss of insufflation occurs. However, the fixtures from the Loop Fixture II device provide enough traction to keep the luminal space from collapsing. Finally, the luminal defect is closed using the Overstitch endoscopic suturing device (Fig. 20).

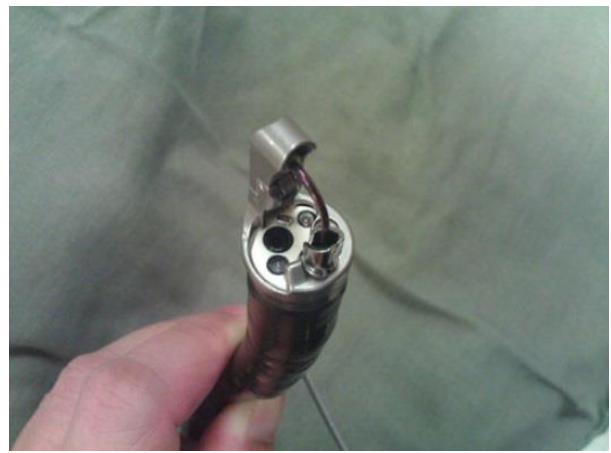

Fig 20: Apollo Overstitch system mounted on a dual channel endoscope

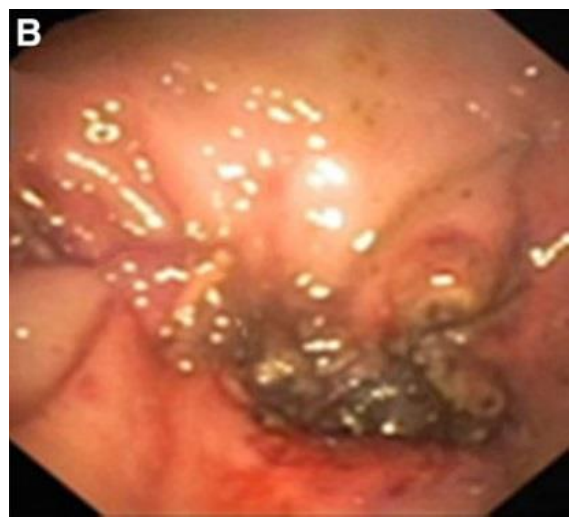

Fig 21: Closure of gastric luminal defect after Overstitch suturing

## Results of Preclinical Trials

Table 3: Procedure time for MASTER full thickness resection

|  | Case 1 | Case 2 |
|---|---|---|
| Total procedural time | 56 min | 70 min |
| Operative time for full-thickness resection using MASTER | 44 min | 52 min |
| Operative time for closure of gastrotomy using Overstitch | 12 min | 18 min |
| Full gastric distension after the procedure | Yes | Yes |
| No. of overstitches applied to close the gastric luminal defect | 2 | 1 |
| Complications during the procedure | Nil | Diathermy injury to anterior abdominal wall |

Preclinical trials were conducted on two nonsurvival porcine specimens with weights of 30 and 35 kg [37]. The average dimensions of the specimens removed was 50 mm by 20 mm. Successful closure of the defects was achieved with satisfactory gastric distension, evidenced by the absence of gas leakage afterwards (See Fig. 21). No injury to adjacent organs was observed. All results are given in Table 3.

# NATURAL ORIFICE TRANSLUMINAL ENDOSCOPIC SURGERY WITH MASTER ROBOT

Natural Orifice Transluminal Endoscopic Surgery (NOTES) refers to a class of procedures in which an endoscope is passed through a natural orifice and then through an incision in the stomach, colon, vagina, or bladder. This allows surgery to be performed in the abdomen without external scars. It has been touted as the next evolution in minimally invasive surgery after laparoscopic surgery. However, a new class of surgical instrument has to be developed before NOTES becomes feasible for widespread use, as the current instrumentation adapted from laparoscopic surgery possess inadequate dexterity and are non-ergonomic.

Hepatic wedge resection is one such procedure in which access can be readily obtained through the gastric wall. Essentially, it is the dissection of the liver to remove tumours through a hole created in the stomach wall. The liver is classified into 8 functional segments according to the blood supply from the hepatic artery and portal vein. Large metastases increase the likelihood that anatomic amounts, i.e. entire segments will have to be removed. Thus, bulk tissue manipulation and effective bleeding control are required during the surgery. Means of closing the gastrotomy and providing insufflation for the peritoneum and stomach are also key requirements. Pre-clinical investigations have been conducted to determine the suitability of the MASTER platform in conducting hepatic wedge resection [28].

## Procedure

Subjects are deprived of food for a period of 18 hours prior and sedated immediately before the surgery. Intubation is performed with an endotracheal tube and general anaesthesia administered. Throughout the surgery, oxygen is supplied through a ventilator. Heart rate and oxygen saturation are monitored every 20 minutes.

A sterile overtube is advanced into the oesophagus with a standard gastroscope. The stomach is irrigated with 10% povidone-iodine antibacterial solution and normal saline to clear the cavity of effluent. The gastroscope is then withdrawn and a dual channel endoscope bearing the MASTER slave arms is inserted. The monopolar cautery hook makes a 10 mm linear incision on the anterior wall of the stomach (See Fig. 22), about 15-20 cm from the gastroesophageal junction. The endoscope then passes through the incision into the peritoneum. The endoscope is flexed to achieve optimal visual registration of the liver. Once the segment to be dissected has been identified, the robotic arms are advanced towards it. The grasper elevates and secures the segment, allowing the cautery hook to dissect it in the appropriate plane. Haemostasis of the cut edges is achieved with the cautery hook. The excised liver segment and the endoscope are then retracted through the gastrotomy and out of the mouth. The gastrotomy can then be closed through suturing or endoclipping.

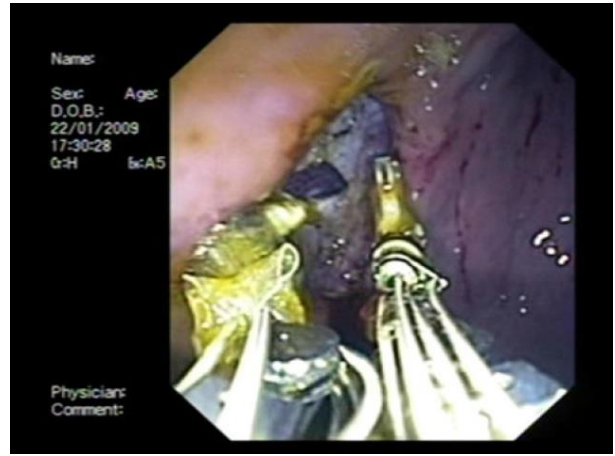

Fig 22: Performance of gastrotomy on the anterior wall of the stomach

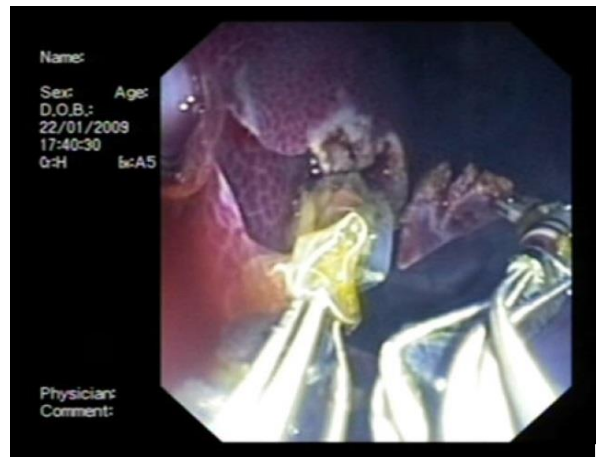

Fig 23: Dissection of the liver segment

## Results of Preclinical Trials

Two porcine subjects successfully underwent natural orifice translumenal hepatic wedge resection using the MASTER.

Table 4: Procedure time for MASTER NOTES surgery

|  | **Case 1** | **Case 2** |
|---|---|---|
| Total procedural time | 10.2 min | 8.5 min |
| Operative time for gastrotomy and approach to liver segment | 2.0 min | 2.5 min |
| Operative time for electrosurgical excision | 8.2 min | 6.0 min |
| Sufficient grasping tension applied to tissue | Yes | Yes |
| Fewer than three attempts at cutting through the liver with monopolar hook at 80W | Yes | Yes |
| Complications during the procedure | Nil | Nil |

Tissue segments of 14 by 8 by 5 mm and 21 by 10 by 7.6 mm were excised and retrieved en bloc. After the operation, the animals were euthanised without closure of the gastrotomy.

# ASSESSMENT OF THE MASTER ROBOTIC ENDOSCOPY SYSTEM

ESD is a difficult procedure because of the lack of a dexterous and ergonomic platform for performing the dissection. Unlike laparoscopy, in which bimanual operation and triangulation of the instruments are possible through different access ports, therapeutic endoscopes make use of a single arm mounted co-axially with the endoscope to manipulate tissue. The high dexterity twin instrumentation of the MASTER robot helps to overcome these limitations. Triangulation is achieved through the bifurcation of the arms at their proximal base joints and re-convergence at the distal joints. Motorised actuation overcomes the tendon friction accumulated along the tortuous endoscope path. An ergonomic human machine interface enables intuitive mapping of user-to-slave motion and greatly reduces the learning curve.

A study was conducted to determine the effect that the MASTER robot had on the learning curve for ESD [39]. Three expert ESD clinicians, three Non-expert ESD clinicians, and three novice non-clinicians were recruited to perform ESD on an ex vivo porcine model using the MASTER. The expert clinicians had each performed more than 100 ESD procedures. The non-expert clinicians had performed less than 10 each, while the novices were engineers who had no prior experience performing the procedure. The novices were able to understand and carry out the ESD procedure within 20 minutes, showing that the learning curve could be significantly shortened and procedure times reduced. Compared with years of training needed for endoscopists to perform ESD, an intensive 1-2 week training duration is sufficient to perform it using the MASTER.

The second generation of the MASTER robot features fully interchangeable instruments that access the surgical site through a conventional endoscope. This obviates the need to swap out the MASTER endoscope for injection of the elevation fluid and allows the full complement of surgical endoscopic tools to be used, such as biopsy forceps, injection needles, snares, and coagulation devices. It is expected to improve the ease of operation further and enable a wide variety of procedures such as full thickness resection and NOTES to be performed reliably.

# FUTURE OF ROBOT-ASSISTED ENDOSCOPIC SURGERY

Robotic therapeutic endoscopes capable of a variety of procedures from GI surface dissection to transendoluminal peritoneal surgery will become commercially available in the near future. This pace of technological development is driven by clinical needs; in line with changing global demographics, the incidence of GI cancer (an age related disease) is expected to increase by 6% per annum. Aside from GI cancer, other diseases and complications treatable by endoscopy include gastroesophageal reflux disease (GERD) and ulcers in the upper GI, diverticulitis, haemorrhoids, Irritable Bowel Syndrome (IBS), and Crohn`s Disease in the lower GI. Government led screening programs will reveal growths at an early stage before they become symptomatic, and patients can opt for their pre-emptive removal by ESD before they advance beyond the submucosal layer. Technical improvements to the tools and interfaces of robot-assisted endoscopic surgical platforms will enhance the capabilities of surgeons, thus allowing them to undertake more ambitious surgeries and push the envelope, especially in the field of NOTES.

## Enhancements to Tools and Interfaces

Better robotic tools and interfaces are the key to improving the feasibility of NOTES procedures and other gasteroenteric treatments. However, some technical challenges need to be addressed before surgeons can be convinced of its efficacy. These include (i) ensuring adequate dexterity, strength, and size of the slave arms, (ii) precision of the slave arms, (iii) function-specific end effectors (iv) the addition of haptic feedback to the system.

## Dexterity, Strength, and Size

From a design perspective, the strength, dexterity, and miniaturisation of a robot arm are competing parameters. A balance has to be struck when developing a robotic arm for a desired task.

Dexterity of the slave arms is required to enable triangulation of instruments and tissue manipulation tasks such as suturing, grasping, retraction/exposure and traction/counter-traction. The human arm has 7 degrees of freedom of motion (shoulder pitch, yaw, roll, elbow pitch, roll and wrist pitch, yaw). In an endoscopic surgery, the elbow and wrist joints are employed the most by the surgeon as large scale relocation is achieved by movement of the entire endoscope. The grasping motion of the fingers also adds another degree of freedom. Hence, at least four DOFs are required for the slave arms, five DOFs if grasping of tissue is required, in order to adequately replicate the configuration of the surgeon's hand and wrist.

Miniaturisation of tools is synonymous with minimally invasive surgery. Current therapeutic dual channel endoscopes are about 12.8 mm in diameter, limiting the possible size of instruments. This in turn limits the raw strength available for manipulation of tissue and forward cutting traction (push or pull) on electrocautery devices. Furthermore, due to friction losses throughout the flexible sheath, the actual force required by the cables at the proximal end could be as high as 100 N (emphasising the need for robotics to assist in applying these loads). Possible alternatives like distal tip mounted actuators are not powerful enough for effective tissue manipulation, but could prove useful for locking or trigger mechanisms.

## Precision

The precision of minimally invasive tools, defined as the robotic arms following the user-desired path as closely as possible, is paramount to ensuring patient safety and operation success. In addition, the accurate control of flexible endoscope also minimizes the error. Precision can help prevent accidental perforations in ESD and enable the efficient application of hemostasis cautery. Precision can be improved by stiffening the endoscope near the surgical site, i.e. it must be flexible enough to navigate through the GI tract to the site, yet stiff enough to prevent flopping of the arms when forces are being exerted on tissue. The path following overtubes used in Medrobotic's Flex is one such innovation. Position sensing feedback and stick-slip friction modelling are also needed to compensate for the jerky motion of cable controlled arms. This can be achieved using the miniaturisation of sensors, which is currently a field of burgeoning scientific interest. Examples include Fibre-Bragg grating sensors that consist of a fibre optic cable running along the length of the endoscope or stretchable MEMS sensors; these can be used to detect its flexed shape. Better spatial cues are also necessary for surgeons, as the narrow angle of image feed from an endoscope can cause surgeons to be disorientated. Studies have shown that in laparoscopic surgeries this is a significant factor that causes damage to vital structures in an operation [40-44].

## End Effectors, Imaging Modalities, Auxiliary Instruments

Despite the encouraging trials performed with the MASTER, more effective instrumentation is needed to make it robust enough for clinical use. More powerful electrosurgery instruments would provide additional headroom to manage any unforeseen major bleeding. More controlled ways of supporting the stomach walls after the loss of insufflation due to gastrotomy could also be developed.

The use of interchangeable instruments, a concept present in normal endoscopes, would allow function-specific end effectors to be inserted for the appropriate task. For example, high payload arms could be used for the manipulation of bulk tissue, while closure of the gastrotomy could be performed by high dexterity arms or specialised suturing tools (similar to the Overstitch endoscopic attachment).

Another innovation will be the development of in vivo imaging modalities that allow one-stop diagnosis and treatment of pre-cancerous tissue. Techniques like Raman spectroscopy and multimodal imaging will allow near instantaneous histopathological assessments at the biomolecular level [45].

## Haptic Feedback

Haptic feedback refers to the somatosensory and proprioceptive stimulation of the user by the machine as a means of conveying information [46-49]. The controllers used in advanced master-slave surgical robots have built-in electronic sensors to detect the relative positions of the human wrist and fingers. They then employ force actuators to adjust the ease at which a user can move his interfacing appendages around in space; this makes it possible to feedback to the user a range of conditions being experienced by the slave robot. For example, a resistance-free sensation is felt by the user when the robot is moving through air but this is dynamically altered into a mushy, damped feeling when contact is made with soft tissue.

The use of haptics improves the ergonomics of using the machine, and crucially, increases patient safety. The absence of haptic feedback on the Da Vinci machine has often been cited as a contributory factor towards surgeon error. With feedback to the surgeon, accidental damage to delicate tissues can be avoided. Software constraints can also be implemented to prevent the over exertion of force. However, challenges remain in integrating force sensors into the small distal tips of the robotic slave arms. The miniaturisation of force and pressure sensors should improve this aspect of surgical robots in the near to mid future.

## Innovations in Surgical Procedures

In the long term, minimally invasive surgery will continue to gain popularity due to the reduction in surgical trauma, which offers better recovery and cosmetic results. Unpredictable factors like antibiotic resistance may also hasten the transition to minimally invasive approaches, as they offer a lower risk of infection than open surgery [50, 51]. Natural orifice transluminal endoscopic surgery is a promising and novel field of surgery that stands to benefit from this evolution. The GI tract runs along the length of the body, offering access to most areas of the peritoneum. Many surgeries could potentially be converted from laparoscopic or open methods to NOTES. For gastroenterologists, collaborations with other specialties will become necessary in order to develop innovative natural orifice surgical approaches which take advantage of the capabilities of robotic platforms.

Currently, robot-assisted endoscopic instruments do not possess the stability, strength, and precision to perform advanced maneuvers on a risk free and reliable basis. These technical issues are the focus of many research groups and surgical technology companies, but in the meantime, one can expect to see hybrid procedures that leverage on the individual strengths of endoscopic and laparoscopic approaches. For example, opening and closure of a gastrotomy can be done using laparoscopic instruments with a peritoneal approach, while a cholecystectomy can be endoscopically performed through the gastrotomy and tissue removed through the mouth [52]. Capsule endoscopy is also a good approach for pre-diagnostic before surgery [53-57].

Eventually, advances in technology will place intuitive, multi-role therapeutic endoscopes in the hands of surgeons. These would allow complex surgical procedures like ESD and NOTES to be performed by surgeons with less experience or skill, thus benefitting a wider population of patients. The technology will also enable more difficult and innovative procedures to be attempted by accomplished surgeons who wish to differentiate themselves and push the boundaries of what is medically possible.